\documentclass[aps,twocolumn,superscriptaddress,showpacs,prl]{revtex4}
\usepackage{bm}
\usepackage{amssymb}
\usepackage{graphicx,epsfig}
\usepackage{subfigure}

\begin{document}
\title{Entanglement and Bell States in Superconducting Flux Qubits}
\author{Mun Dae \surname{Kim}}
\email{mdkim@kias.re.kr}
\affiliation{Korea Institute for Advanced Study, Seoul 130-722, Korea}
\author{Sam Young \surname{Cho}}
\email{sycho@physics.uq.edu.au}
\affiliation{Department of Physics, Chongqing University, Chongqing
400044, The People's Republic of China}
\affiliation{Department of Physics, The University of Queensland, 4072,
Australia}

\begin{abstract}
We theoretically study macroscopic quantum entanglement
in two superconducting flux qubits.
To manipulate the state of two flux qubits,
a Josephson junction is introduced in the connecting loop coupling the qubits.
Increasing the coupling energy of the Josephson junction makes it possible
to achieve relatively strong coupling between the qubits,
causing two-qubit tunneling processes to be even dominant over
the single-qubit tunneling processes in the states of two qubits.
It is shown that due to the two-qubit tunneling processes
both the ground state and excited states of the coupled flux qubits can be
a Bell type state, maximally entangled, in experimentally accessible regimes.
The parameter regimes for the Bell states are
discussed in terms of magnetic flux and Josephson coupling energies.
\end{abstract}

\maketitle

{\it Introduction.}$-$
Quantum entanglement is a fundamental resource
for quantum information processing and quantum computing \cite{Nielsen}.
Numerous proposals have been made for the creation of entanglement
in solid-state systems: quantum dots \cite{Glazman,Marcus,Engel,Piermarocchi},
Kondo impurities \cite{Cho}, carbon nanotubes \cite{Ardavan} and so on.
For superconducting qubits
coherent manipulation of quantum states in a controllable manner
has enabled to generate  partially entangled states
\cite{Pashkin, Berkley, Izmalkov}.
However, in realizing such quantum technologies,
highly entangled quantum states such as the Bell states of two qubits
are required \cite{Nielsen}. Of particular importance,
therefore, are the preparation and measurement of such maximally
entangled states.

Recently, several types of superconducting qubits have been demonstrated
experimentally.
In superconducting qubits experimental generations of  entanglements
have been reported for coupled charge qubits \cite{Pashkin}
and coupled phase qubits \cite{Berkley},
but the maximally entangled Bell states is far from
experimental realization as yet.
This paper aims to answer on how maximally entangled states
can be prepared by manipulating system parameters in solid-state qubits.
We quantify quantum entanglement in the two superconducting flux qubits
based on the phase-coupling scheme.
In this study
we show that,
if the coupling strength between two  flux qubits is
strong enough,
simultaneous {\it two-qubit coherent tunneling} processes
make it possible to create a maximally entangled state
in the ground and excited states.
Furthermore, it is shown that
the ranges of system parameters for a maximally entangled state
are sufficiently wide that
a Bell-type state should be realizable experimentally.
Actually, a coherent two-qubit flipping processes
has been experimentally observed
in inductively coupled flux qubits \cite{Izmalkov}.
However, the strength of the inductive coupling \cite{Majer}  is too weak
to achieve a maximally entangled state.

\begin{figure}[b]
\vspace{2.7cm}
\includegraphics{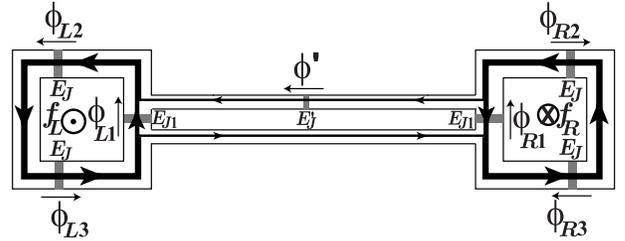}
\vspace*{0cm}
\caption{
 Left and right superconducting loops are
 connected each other by a connecting loop interrupted by a Josephson junction.
 The state of each qubit loop is the superposed state of
 the diamagnetic and paramagnetic current states assigned by
 $\left|\downarrow\right\rangle$ and  $\left|\uparrow\right\rangle$,
 respectively, which can make the loop being regarded as a qubit.
 For example,
 the state, $\left|\downarrow\uparrow\right\rangle$,
 out of four possible basis of two-qubit current states
 is shown, where the arrows indicate the flow of Cooper pairs and
 thus in reverse direction is the current.
 Here, $\odot$ (oppositely $\otimes$) denote the directions of
 the magnetic fields, $f_{L(R)}=\Phi_{L(R)}/\Phi_0$, in the qubit loops.
 $E_{J1}$, $E_J$, and $E'_J$ are
 the Josephson coupling energies of the Josephson junctions in the qubit loops
and the connecting loop and $\phi$'s are phase differences across the Josephson junctions.
 }
\label{Coup3JJs}
\end{figure}

We use a phase-coupling scheme \cite{Kim}
to obtain sufficiently strong coupling
and show a maximally entangled state between two flux qubits.
Very recently, this phase-coupling scheme
has been realized in an experiment \cite{Ploeg}
using  two four-junctions flux qubits.
 To theoretically study a controllable coupling manner
 in the flux qubits, the phase coupling scheme has
 also been employed \cite{KimCC,Grajcar}.
 Further, there have been studies about somewhat
 different types of phase-coupling schemes
\cite{Grajcar3, Brink}.
The phase-coupling scheme for two flux qubits (See Fig. \ref{Coup3JJs})
is to introduce a connecting loop interrupted by a Josephson junction
in order to couple two three-junctions qubits \cite{Mooij,Kim1}.
In the connecting loop, the Josephson energy
depends on the phase difference, $\phi_{L1}-\phi_{R1}$,
and the coupling energy, $E'_J$, of the junction.
As varies $E'_J$,
the coupling strength between the flux qubits, defined by the energy difference
between the same direction current state and the different
direction current state in the two flux qubits,
can be increased to be relatively strong.
The coupling strength between the phase-coupled flux qubits
is a monotonously increasing function of $E'_J$ \cite{Kim}.
For small $E_J'(\ll E_J)$,
the effective potential for the two qubits has a symmetric form in
the phase variable space.
Then single-qubit tunneling processes in the two qubit states are predominant
in the effective potential.
They have almost the same tunneling amplitudes.
The eigenstates are  nearly degenerate
at the operating point of the external fluxes
so that the entanglement between two qubits is very weak.
As increases the Josephson coupling energy, $E_J'$,
 the shape of the effective potential
 is deformed to be a less symmetric form.
 Then the amplitudes of single-qubit tunneling processes
 are decreased.
 However, the deformation of the effective potential
 allows a two-qubit tunneling process not negligible
 and even dominant over the single-qubit tunneling processes.
 We will then
 derive explicitly the contribution of two-qubit tunneling processes.
 To generate a Bell type of maximally entangled
state, the two-qubit tunneling processes are shown to play an important role.

{\it Model.}$-$
Let us start with
the charging energy of the Josephson junctions:
$E_C=\frac12 \left(\Phi_0/2\pi\right)^2(
\sum^3_{i=1} (C_{Li}\dot{\phi}^2_{Li}+C_{Ri}\dot{\phi}^2_{Ri})
+C'\dot{\phi'}^2 )$,
where $C_{L(R)i}$ and $C'$ are the capacitance of the Josephson junctions
for the left (right) qubit loop and the connecting loop, respectively.
Here, $\phi$'s are the phase differences across the Josephson junctions
and  $\Phi_0\equiv h/2e$ the superconducting unit flux quantum.
Since the number of excess Cooper pair charges on the Josephson junctions,
$\hat{N_i}\equiv \hat{Q}_i/2e$ with $Q_i=C(\Phi_0/2\pi){\dot\phi_i}$,
is conjugate to the phase difference $\hat{\phi}_i$
such as $[\hat{\phi}_i,\hat{N_i}]=i$,
the canonical momentum,
$\hat{P}_i\equiv \hat{N_i}\hbar= -i\hbar\partial /\partial \hat{\phi}_i$,
can be introduced.
Then the Hamiltonian is given by
\begin{eqnarray}
\label{Hamiltonian}
\hat{H}=\frac12\hat{ P}^T_i{M}^{-1}_{ij}\hat{ P}_j+U_{\rm eff}(\hat{\bm{\phi}}),
\end{eqnarray}
where
$M_{ij}=(\Phi_0/2\pi)^2C_i\delta_{ij}$
is the effective mass and
$\hat{\bm{\phi}}=(\phi_{L1}, \phi_{L2}, \phi_{L3}, \phi_{R1},
\phi_{R2}, \phi_{R3}, \phi')$.
If we neglect the small inductive energy,
the effective potential
becomes the energy of the Josephson junctions such as
$U_{\rm eff}(\bm{\phi})=\sum^3_{i=1}E_{Ji}(1-\cos\phi_{Li})
+\sum^3_{i=1}E_{Ji}(1-\cos\phi_{Ri})+E'_{J}(1-\cos\phi')$.

\begin{figure}[t]
\vspace*{6.4cm}
\includegraphics{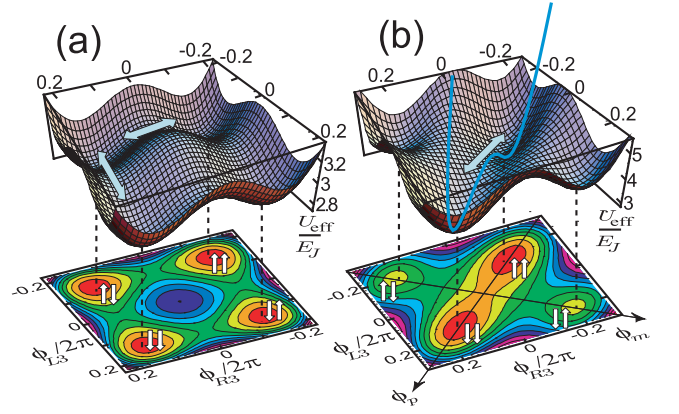}
\vspace*{-1cm}
\caption{(Color online)
 The effective potential of Eq. (\ref{Ueff}) for the coupled flux qubits
 at the co-resonance point $f_L=f_R=0.5$
 as a function of $\phi_{L3}$ and $\phi_{R3}$
with $E_{J1}=0.7 E_J$.
 In the contour plots of the effective potential, the four local minima
 correspond to the current states of the coupled flux qubits:
          $\left|\uparrow\uparrow \right\rangle,
           \left|\uparrow\downarrow \right\rangle,
           \left|\downarrow\uparrow \right\rangle$,
           and $\left|\downarrow\downarrow \right\rangle$.
 The left-right arrows ($\Longleftrightarrow$)
 on the effective potential profiles indicate
 some of single-/two-qubit tunneling processes
 (For example,
  $\left|\downarrow\uparrow \right\rangle  \Longleftrightarrow
          \left|\uparrow\uparrow \right\rangle $
 and
  $\left|\downarrow\downarrow \right\rangle  \Longleftrightarrow
          \left|\uparrow\uparrow \right\rangle $, respectively).
 (a) For $E'_J=0$,
  the four local minima of the effective potential have
  nearly the same energies.
  The potential barrier of the double well for a two-qubit tunneling process
  are much wider and higher than the potential barriers for the single-qubit
  tunneling processes.
 Thus, single-qubit tunneling processes are dominant and the entanglement is very weak.
 (b) For $E'_J=0.6E_J$,
 only $\left|\downarrow\uparrow\right\rangle$  and
 $\left|\uparrow\downarrow\right\rangle$
 are lifted up to a higher energy while
 $\left|\uparrow\uparrow\right\rangle$  and
 $\left|\downarrow\downarrow\right\rangle$
 stay at the same energies.
 Then the potential barrier for the two-qubit tunneling process between
 $\left|\downarrow\downarrow\right\rangle$ and
 $\left|\uparrow\uparrow\right\rangle$
 is not changed so much that the tunneling amplitude also is not changed.
 An asymmetric double well potential for single-qubit tunneling processes
 makes the tunneling amplitude decreasing as $E_J'$ increases.
 Therefore, the two-qubit tunneling process plays a significant role
 on the entanglement of the coupled flux qubits.
}
\label{Potential}
\end{figure}

Since the qubit operations are performed experimentally
at near the co-resonance point, we can set $f_L=f_R=0.5$ with $f_{L(R)}\equiv \Phi_{L(R)}/\Phi_0$
and the total flux $\Phi_{L(R)}$ threading the left (right) qubit loop.
In experiments,
the two Josephson junctions with phase differences, $\phi_{L(R)2}$
and $\phi_{L(R)3}$, are considered nominally the same
so that it is reasonable to set $E_{J2}=E_{J3}=E_J$ and
$\phi_{L(R)2}=\phi_{L(R)3}$.
Furthermore, if one can neglect the small inductive flux,
the boundary conditions in the left and right qubit loops
and the connecting loop are given approximately,
$2\pi (n_{L(R)}+ f_{L(R)})-(\phi_{L(R)1}+\phi_{L(R)2}+\phi_{L(R)3})=0$
and $2\pi r+(\phi_{L1}-\phi_{R1})-\phi'=0$
with integers $n_L, n_R$ and $r$.
Introducing the rotated coordinates with
$\phi_p\equiv (\phi_{L3}+\phi_{R3})/2$
and $\phi_m\equiv (\phi_{L3}-\phi_{R3})/2$,
the Hamiltonian can be rewritten in the form,
${\hat H}={\hat P}_p^2/2M_p+{\hat P}_m^2/2M_m+U_{\rm eff}
(\hat{\phi}_p,\hat{\phi}_m),$
where the effective potential becomes
\begin{eqnarray}
\label{Ueff}
U_{\rm eff}(\phi_p,\phi_m)=2E_{J1}(1+\cos2\phi_p\cos2\phi_m)~~~~~~~~~~ \\
+4E_J(1-\cos\phi_p\cos\phi_m)+E'_J(1-\cos4\phi_m). \nonumber
\end{eqnarray}
Here,
$M_p\equiv \hbar^2 (2C_1+C)/e^2$,
$M_m\equiv \hbar^2 (2C_1+C+4C')/e^2$ and the conjugate momentum
is $\hat{P}_{p(m)}=-i\hbar\partial/\partial\hat{\phi}_{p(m)}$
with commutation relation, $[\hat{\phi}_{p(m)},\hat{P}_{p(m)}]=i\hbar$.

 We display the effective potential as a function of
 $\phi_{L3}$ and $\phi_{R3}$ in Fig. \ref{Potential}.
 The effective potential is shown to have the four local potential minima
 corresponding to the four states of the coupled qubits, i.e.,
          $\left|\uparrow\uparrow \right\rangle,
           \left|\uparrow\downarrow \right\rangle,
           \left|\downarrow\uparrow \right\rangle$,
           and $\left|\downarrow\downarrow \right\rangle$.
At the local minima of the effective potential,
one can obtain the energy levels:
$E_{ss'}=(1/2)\hbar(\omega_{p,ss'}+\omega_{m,ss'})
+U_{\rm eff}(\phi_{p,ss'},\phi_{m,ss'})$,
where $s,s'=1/2$ and  $-1/2$ stand for spin up, $\left|\uparrow\right\rangle$,
and spin down, $\left|\downarrow\right\rangle$, respectively.
In the harmonic oscillator approximation \cite{Orlando},
the characteristic oscillating frequencies, $\omega_{p(m),ss'}$,
are given by
$\omega_{p(m),ss'}=(k_{p(m),ss'}/M_{p(m)})^{1/2}$  with
$k_{p(m),ss'}\equiv \partial^2U_{\rm eff}
(\phi_p,\phi_m)/\partial\phi^2_{p(m)}|_{(\phi_{p,ss'},\phi_{m,ss'})}$,
where $\phi_{p,ss'} (\phi_{m,ss'})$ are
the values of $\phi_{p} (\phi_{m})$ at the local potential minimum.
Then the tight-binding approximation
gives the Hamiltonian in the basis,
          $\{ \left|\uparrow\uparrow \right\rangle,
           \left|\uparrow\downarrow \right\rangle,
           \left|\downarrow\uparrow \right\rangle,
           \left|\downarrow\downarrow \right\rangle \}$,
as follows;
\begin{eqnarray}
\label{totalH}
H\!\! &=&\!\!\!\!\!\! \sum_{s,s'=\pm \frac12}
    E_{ss'}|s,s'\rangle\langle s,s'|+H^T_1+ H^T_2,\\
\label{H1}
H^T_1 \!\!&=& \!\!\!\!\!\! \sum_{s,s'=\pm \frac12 }\!\!
   \Big(-t^L_1|s,s'\rangle\langle -s,s'|-t^R_1|s,s'\rangle\langle s,-s'|\Big),\\
\label{H2}
H^T_2 \!\!&=& \!\!\! \sum_{s=\pm \frac12 }
    \Big(-t^a_2|s,s\rangle\langle -s,-s|-t^b_2|s,-s\rangle\langle -s,s|\Big),
\end{eqnarray}
where $H^T_{1(2)}$ describe the single(two)-qubit tunneling processes between
the two-qubit states with the tunneling amplitudes $t^{L,R}_1$($t^{a,b}_2$).
Normally, the single-qubit tunneling amplitudes are much larger than
the two-qubit tunneling amplitudes, i.e., $t^{a,b}_2 \ll t^{L,R}_1$.
Thus, the two-qubit tunneling Hamiltonian $H^T_2$ can be neglected
when $E_J'$=0.

 {\it Two-qubit tunnelings and entanglements.}$-$
For the weak coupling limit, $E_J' \ll E_J$,
the single-qubit tunnelings between the two-qubit states
can be obtained by the tunnelings in
the well-behaved double well potentials
as shown as left-right arrows in Fig. \ref{Potential}(a).
As increases $E_J'$,
the well-behaved double well potentials for the single-qubit tunnelings
become an asymmetric double well potentials
shown clearly in Fig. \ref{Potential}(b).
Then the asymmetry of the double well potentials
makes the single-qubit tunneling amplitudes decrease drastically.
However,
as the blue line shown in Fig. \ref{Potential}(b),
the corresponding double well potential to
the two-qubit tunneling between the states
$\left|\uparrow\uparrow\right\rangle$ and
$\left|\downarrow\downarrow\right\rangle$
is not changed much qualitatively.
Actually, from the effective potential in Eq. (\ref{Ueff}),
the double well potential is given by
\begin{eqnarray}
\label{well}
U_{\rm eff}(\phi_p,0)=2E_{J1}(1+\cos2\phi_p)+4E_J(1-\cos\phi_p).
\end{eqnarray}
The two-qubit tunneling amplitudes $t^a_2$ between the states
$\left|\uparrow\uparrow\right\rangle$ and
$\left|\downarrow\downarrow\right\rangle$ is written in the WKB approximation \cite{Orlando},
\begin{eqnarray}
\label{t2a}
\!\!\!\!\!\!\!t^{a}_{2}\!\!\! &=&\!\!\! \frac{\hbar\omega_{p,ss}}{2\pi}
 \exp \!\! {\left[\!-\sqrt{\frac{2M_p}{\hbar^2}}\!
    \int \!\! d\phi_p\sqrt{U_{\rm eff}(\phi_p,0)-E_{ss}}\right]}.
\end{eqnarray}
Here $M_p$, $\omega_{p,ss}$ and $U_{\rm eff}(\phi_p,0)$
do not depend on the capacitance, $C'$, and
the Josephson coupling energy, $E'_J$, of the Josephson junction in
the connecting loop.
Equation (\ref{t2a}) shows that
the two-qubit tunneling amplitude remains unchanged
even though $E'_J$ is increased.
As a consequence, the two-qubit tunneling between
the states $\left|\uparrow\uparrow\right\rangle$ and
$\left|\downarrow\downarrow\right\rangle$ in Hamiltonian $H^T_2$
of Eq. (\ref{H2}) will play a crucial role to improve the entanglement between
the coupled flux qubits.

For $E_J'=0$,
we obtain the tunneling amplitudes;
$t^{L,R}_1\approx 0.0075E_J$ and $t^{a}_2=t^{b}_2\approx 0.00024E_J$ with $E_{J1}=0.7E_J$.
As expected,
$t^{a,b}_2 \ll t^{L,R}_1$ and the two-qubit tunneling terms
of the Hamiltonian are negligible.
When $E'_J\neq 0$,
in order to get the single-qubit tunneling amplitudes
between an asymmetric double well potentials,
the Fourier Grid Hamiltonian Method \cite{Marton}
is employed because the WKB approximation cannot be applicable
in the asymmetric double well potential.
For $E_J'=0.6E_J$ as shown in Fig. \ref{Potential}(b),
we obtain the  tunneling amplitudes
of $t^{L,R}_{1}\approx 10^{-5} E_J$
and $t^{a}_2\approx 0.00024E_J$.
Here,
the tunneling amplitude $t^b_2$ is neglected because
the the wave function overlap between the states
 $\left|\uparrow\downarrow\right\rangle$ and
 $\left|\downarrow\uparrow\right\rangle$ is negligible
in Fig. \ref{Potential}(b).

\begin{figure}[t]
\vspace*{4cm}
\includegraphics{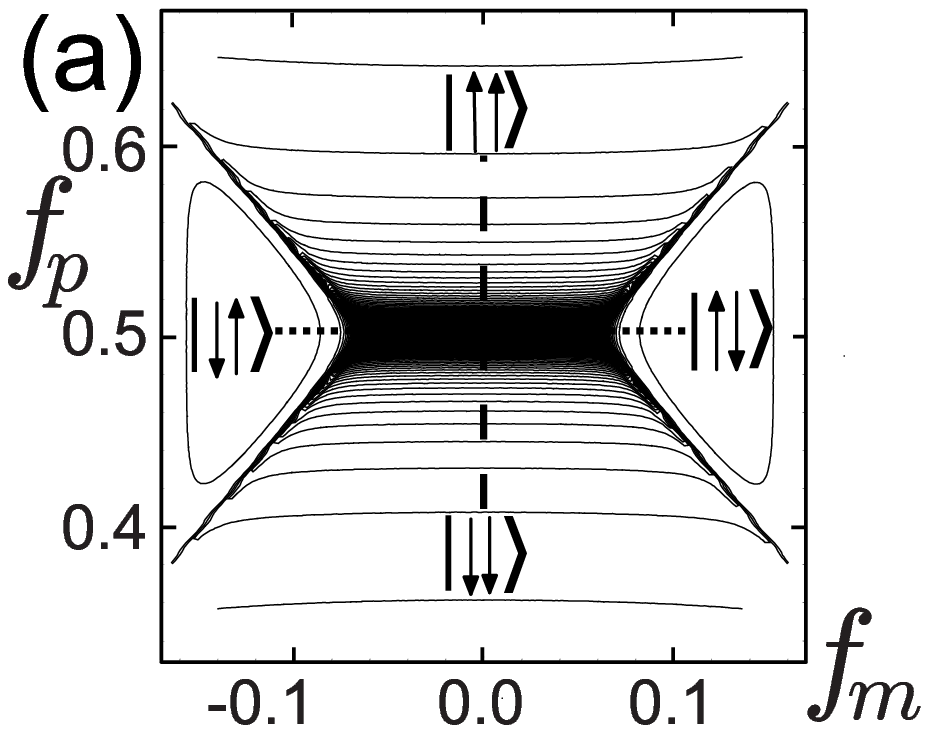}
\vspace{3cm}
\includegraphics{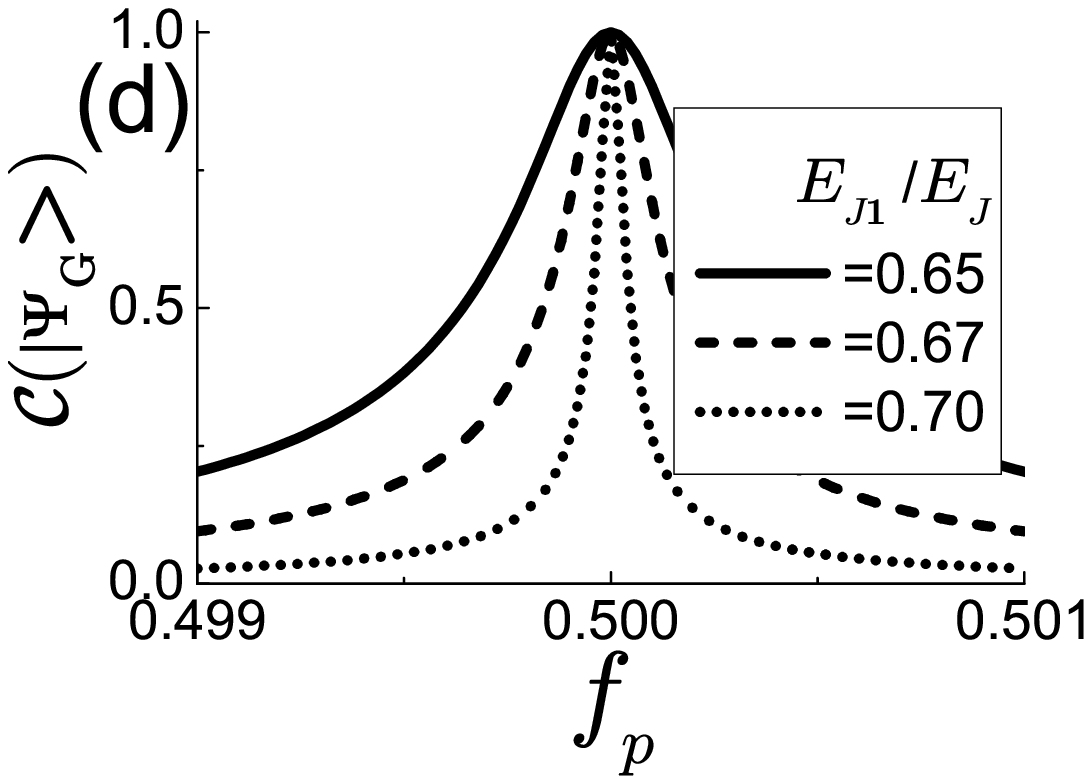}
\hspace{0cm}
\includegraphics{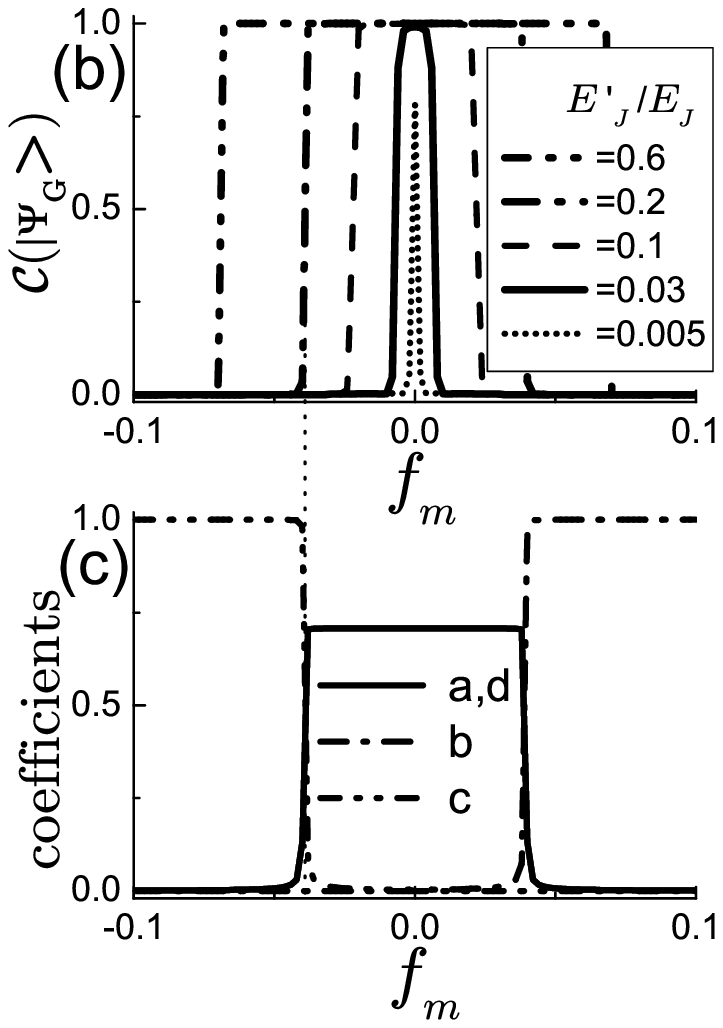}
\vspace{-0.5cm}
\caption{(a) Concurrence for the ground state of the coupled flux qubits
         as a function of $f_{p(m)}=(f_L\pm f_R)/2$
     for $E'_J=0.6E_J$ and $E_{J1}=0.7E_J$.
     A high entanglement between two flux qubits is seen
     in the broad ridge with $f_p=0.5$.
    In the contour plot,  $\left|\downarrow\downarrow\right\rangle$,
    $\left|\uparrow\downarrow\right\rangle$,
    $\left|\downarrow\uparrow\right\rangle$,  and
    $\left|\uparrow\uparrow\right\rangle$
   denote  the coupled-qubit states located
   far from the co-resonance point, $f_L=f_R=0.5$.
 (b) Concurrences for various $E'_J$ along the line of $f_p=0.5$ (dotted line
     in (a)) for $E_{J1}=0.7E_J$. As increases  $E'_J$,
     the maximum entanglement appears
     for $E'_J \gtrsim  0.03\, E_J$.
 (c) The coefficients  of the ground state wavefunction,
  $\left|\psi_G\right\rangle=a(f_m) \left|\downarrow\downarrow\right\rangle
   +b(f_m) \left|\uparrow\downarrow\right\rangle
   +c(f_m) \left|\downarrow\uparrow\right\rangle
   +d(f_m) \left|\uparrow\uparrow\right\rangle$,
  as a function of $f_m$  for the case $E'_J=0.2E_J$ in (b).
 For $-0.04 \lesssim f_m \lesssim 0.04$,
 the Bell type of maximally entangled state,
  $|\Phi^+\rangle=(\left|\downarrow\downarrow\right\rangle
            +\left|\uparrow\uparrow\right\rangle)/\sqrt{2}$,
  is shown clearly because $a=d=1/\sqrt{2}$  and $b = c =0$.
 (d)
 Concurrences  for various $E_{J1}$
 along the line of $f_m=0$
 for $E'_J=0.6E_J$ (dashed line in (a)).
 As decreases $E_{J1}$, since the two-qubit tunneling amplitude increases,
 the width of concurrence peak becomes so wide
 that the maximal entanglement
 should be observable experimentally.
 }
\label{Conc}
\end{figure}

One measure of entanglement is  the concurrence, ${\cal C}$,
for an arbitrary state of coupled two qubits.
The concurrence ranges from 0 for nonentangled to 1
for maximally entangled states.
For a normalized pure state,
$\left|\psi\right\rangle=a \left|\downarrow\downarrow\right\rangle
 +b \left|\uparrow\downarrow\right\rangle
 +c \left|\downarrow\uparrow\right\rangle
 +d \left|\uparrow\uparrow\right\rangle$,
 the concurrence \cite{Wootters} is given by
\begin{equation}
\label{conc}
 {\cal C}(\left|\psi\right\rangle)
 =2\left|ad-bc\right|.
\end{equation}
We evaluate the concurrence in Eq. (\ref{conc}) numerically
to show that a maximally entangled state
is possible in the ground state by varying
the Josephson coupling energy in the connecting loop.
In Fig. \ref{Conc}(a), we plot
the concurrence for the ground state, ${\cal C}(\left|\psi_G\right\rangle)$,
of the Hamiltonian in Eqs. (\ref{totalH})$-$(\ref{H2})
as a function of $f_{p(m)}=(f_L\pm f_R)/2$
for $E_J'=0.6E_J$ and $E_{J1}=0.7E_J$.
A broad ridge of the concurrence
along the line $f_p=0.5$ shows
a high entanglement between the two qubits.
This region corresponds to the central part of the honeycomb type potential of coupled qubits
in Ref. \onlinecite{Kim} near the co-resonance point, $f_L=f_R=0.5$.
Away from the co-resonance point we can numerically calculate concurrences without
using the analytic effective potential in Eq. (\ref{Ueff})
obtained by introducing a few approximations \cite{Kim}.
Figure \ref{Conc}(b) shows the cut view of the concurrence at $f_p=0.5$
(dotted line in Fig. \ref{Conc}(a))
as a function of $f_m$ for various values of $E_J'$.
For $E'_J=0.005E_J$, corresponding to
the coupling strength of inductively coupled qubits in
the experiment of Ref. \onlinecite{Majer},
a partial entanglement can only exist around $f_m \sim 0$.
As increases $E_J'$, i.e., when the two-qubit tunneling
becomes dominant over the single-qubit tunneling,
a maximum entanglement appears.
In Fig. \ref{Conc}(c),
we display
  the coefficients of the ground state wavefunction,
  $\left|\psi_G\right\rangle=a(f_m) \left|\downarrow\downarrow\right\rangle
   +b(f_m) \left|\uparrow\downarrow\right\rangle
   +c(f_m) \left|\downarrow\uparrow\right\rangle
   +d(f_m) \left|\uparrow\uparrow\right\rangle$,
  as a function of $f_m$ for $E'_J=0.2E_J$.
  We see the ground state wavefunction:
\begin{eqnarray}
  \left|\psi_G\right\rangle
  \simeq \! \left\{
    \begin{array}{ll}
    \left|\downarrow\uparrow\right\rangle & \mbox{for $f_m \lesssim -0.04$} \\
    \!\!\! \displaystyle  \frac{1}{\sqrt{2}}
    \left( \left|\downarrow\downarrow\right\rangle
         + \left|\uparrow\uparrow\right\rangle \right)&
         \mbox{for $-0.04 \lesssim f_m \lesssim 0.04$}  \\
     \left|\uparrow\downarrow\right\rangle &
         \mbox{for $f_m \gtrsim 0.04$}  \\
          \end{array} \right.\!\! .
\end{eqnarray}
 This ground state has the Bell type of maximally entangled state,
  $|\Phi^+\rangle=(\left|\downarrow\downarrow\right\rangle
            +\left|\uparrow\uparrow\right\rangle)/\sqrt{2}$,
 for $-0.04 \lesssim f_m \lesssim 0.04$.
 Similarly, one can see that
 the 1st excited state shows, actually, another Bell type
 of maximally entangled state,
 $\left|\Phi^-\right\rangle=
  (\left|\downarrow\downarrow\right\rangle
      -\left|\uparrow\uparrow\right\rangle)/\sqrt{2}$.

The cut view of the concurrence
at $f_m=0$ (dashed line in (a))
is shown in Fig. \ref{Conc}(d)
for various $E_{J1}$ with $E'_J=0.6E_J$.
As $E_{J1}$ increases,
the barrier of the double well potential in Eq. (\ref{well})
becomes higher.
Thus the two-qubit tunneling amplitude, $t^a_2$, decreases such that
$t^a_2/E_J\approx$ 0.0016, 0.0008, 0.00024
for $E_{J1}/E_J$=0.65, 0.67, 0.7, respectively.
 In Fig. \ref{Conc}(d), as a result, the width of the concurrence peak
 becomes narrower,
which will make an experimental implementation of
the maximum entanglement difficult.
 On the other hand, if the value of $E_{J1}$ becomes much smaller
 than $E_{J1}=0.65E_J$, the barrier of double well potential can be
 lower than the ground energy of the harmonic potential wells,
 $E_{ss}$, and the two-qubit  states,
 $\left|\downarrow\downarrow\right\rangle$
 and $\left|\uparrow\uparrow\right\rangle$, in Fig. \ref{Potential}(b)
 will not be stable.
In addition, as seen in Fig. \ref{Conc}(b),
the range of $E'_J$ for maximal entanglement is approximately
$E_J'\gtrsim  0.03 E_J$.
Therefore,
for a maximally entangled state of the two flux qubits,
it  is required that
 $E_{J1}$ should be controlled
 around $0.7E_J$ with $E'_J \gtrsim 0.03 \, E_J$.

For $E'_J \ll E_J$,
the concurrence is very small because
the four states are nearly degenerated due to the weak coupling between
the qubits and $t^{L,R}_1 \gg t^{a,b}_2$.
As $E'_J$ increases, the coupling between the qubits is much stronger
and $t^a_2$ becomes much dominant over than other tunneling
processes, $t^{L}_1$, $t^{R}_1$, and $t^b_2$.
In a strong coupling limit,  $t^{L}_1, t^{R}_1, t^b_2 \ll t^a_2$ and
$E_{s,s} < E_{s,-s}$.
If we
with the split energies,
$E_{\downarrow\downarrow}=E_0-\delta E$
and $E_{\uparrow\uparrow}=E_0+\delta E$
around the co-resonance point.
$f_L=f_R \approx 0.5$
Then the concurrence of the ground state is given by
${\cal C}(\left|\psi_G\right\rangle)
 \simeq | t^a_2/\sqrt{(t^a_2)^2+(\delta E)^2}|$,
which corresponds to the behaviors in Fig. \ref{Conc}(d).
At the co-resonance point ($f_L=f_R=0.5$),
the ground state is in a maximally entangled state, i.e.,
${\cal C}(\left|\psi_G\right\rangle)=1$.
This shows the important role of the two-qubit tunneling process making
the two qubits entangled and maintaining the highly entangled state against
fluctuations of $f_{L,R}$ away from the co-resonance point as long as
$\delta E \lesssim t^a_2$.

{\it Summary.}$-$
 We studied the entanglement to achieve a maximally entangled state
 in a coupled superconducting flux qubits.
 A Josephson junction in the connecting loop
 coupling the two qubits was employed to manipulate the qubit states.
 As increases the Josephson coupling energy of the Josephson junction,
 the two-qubit tunneling processes
 between the  current states
 play an important role to make
 the two flux qubit strongly entangled.
 It was shown that a Bell type of maximally entangled states
 can be realized in the ground and excited state
 of the coupled qubit system.
 We also identified the system parameter regime for the
 maximally entangled states.

 {\it Acknowledgments.}
 We thank Yasunobu Nakamura for helpful discussions.
 This work was supported by the Ministry of Science \&
 Technology of Korea (Quantum Information Science) and
 the Australian Research Council.

\end{document}